\documentclass{article}
\usepackage{hiph-art}
\usepackage{graphicx}
\usepackage{bm}
\usepackage{color}

\volnumber{19} \issuenumber{1} \edyear{2004}                             
\frompage{000} \topage{000}                                              
\recrevdate{21 November 2005}                                            
\def\rightmark{Numerical Simulation of Non-Abelian Particle-Field Dynamics}
\def\leftmark{A.\ Dumitru and Y.\ Nara}
 
\title{Numerical Simulation of Non-Abelian Particle-Field Dynamics}
\authors{ 
{Adrian Dumitru and Yasushi Nara
}\\[2.812mm]
{\normalsize
Institut f\"ur Theoretische Physik,
Johann Wolfgang Goethe Universit\"at,\\
Max von Laue Str.\ 1, D-60438 Frankfurt am Main, Germany
}}
 
\abstract{
Numerical 1D-3V solutions of the Wong-Yang-Mills equations
with anisotropic particle momentum distributions
are presented. They confirm the existence of plasma instabilities for
weak initial fields and of their saturation at a level where the
particle motion is affected, similar to Abelian plasmas.
The isotropization of the particle momenta by strong random fields is shown
explicitly, as well as their nearly exponential distribution up to
a typical hard scale, which arises from scattering off field
fluctuations. By variation of the lattice spacing we show that the
effects described here are independent of the UV field modes near the
end of the Brioullin zone.
}


\makeindex
\begin{document}
 
\maketitle

\section{Introduction}\label{intro}
Recently, it has been realized that non-Abelian collective plasma
processes such as Weibel-like instabilities might play an important
role for the thermalization process in the early stage of high-energy
heavy-ion collisions. This was the central topic of the workshop on
``Quark-Gluon-Plasma Thermalization'' in Vienna~\cite{QGPTH05}.  If
so, a quantitative understanding of such processes will be crucial to
answer, for example, the question about the maximum temperature
achieved in such collisions at the BNL-RHIC and CERN-LHC colliders.

The physics of non-Abelian plasma instabilities in the context of
relativistic heavy-ion collisions has been discussed in some detail in
a recent review~\cite{Stan_review} and in many contributions to this
workshop. We shall therefore refrain from a detailed presentation
here. Rather, we focus on illustrating the generalization of Abelian
particle-in-cell simulations to the SU(2) gauge group. These
provide some additional insight into the physics of non-Abelian
plasmas beyond the ``Hard Loop''
approximation which underlies much of the present analytical and numerical
understanding of SU(2) instabilities~\cite{HL}. Simulations of the
non-linear Vlasov-Yang-Mills theory
account for the back-reaction of the fields on the particles, which
damps (and eventually shuts off) the exponential growth of the
chromo-magnetic fields. They also enable us to actually look at the
time-evolution and eventual isotropization of the particle momenta
themselves. Another motivation for performing full particle-field
simulations is the potential interest in initial conditions where the
fields are strong and immediately affect the particle motion.
Some of our results have been published in ref.~\cite{DN05}.

\section{Particle-in-cell simulations for non-Abelian gauge theories}

We consider the classical transport equation
for hard gluons with non-Abelian color charge $Q^a$ in the
collisionless approximation~\cite{Wong,ClTransp},
\begin{equation}
 p^{\mu}[\partial_\mu - gQ^aF^a_{\mu\nu}\partial^\nu_p
    - gf_{abc}A^b_\mu Q^c\partial_{Q^a}]f(x,p,Q)=0~.
\end{equation}
where $f$ denotes the one-particle phase-space distribution
function. We employ the test-particle method, replacing
the continuous distribution $f(t,\bm{x},\bm{p},\bm{Q})$
by a large number of test particles:
\begin{equation}
 f(t,\bm{x},\bm{p},Q) = \frac{1}{N_{test}}\sum_i 
  \delta(\bm{x}-\bm{x}_i(t)) (2\pi)^3 \delta(\bm{p}-\bm{p}_i(t)) 
         \delta(\bm{Q}-\bm{Q}_i(t))~.
\end{equation}
$\bm{r}_i(t)$, $\bm{p}_i(t)$, $\bm{Q}_i(t)$ are the phase-space coordinates of
an individual test-particle. (We consider particles in the adjoint
representation of color-SU($N_c$), hence $\bm{Q}$ is a vector in
$N_c^2-1$ dimensional color space.)
This {\sl Ansatz} leads to Wong's equations~\cite{Wong,ClTransp}
\begin{equation}
\frac{d\bm{x}_i}{dt}  =  \bm{v}_i,\qquad
\frac{d\bm{p}_i}{dt}  =  gQ_i^a \left( \bm{E}^a + \bm{v}_i \times
  \bm{B}^a \right),\label{pdot}\qquad
\frac{dQ_i}{dt}  =  igv^{\mu}_i [ A_\mu, Q_i],
\end{equation}
for the $i$-th (test) particle, coupled to the Yang-Mills equation
\begin{equation}
\frac{dA^i}{dt}= E^i,\qquad
\frac{dE^i}{dt}=D_jF^{ji} - \frac{g}{N_{\mathrm{test}}}
  \sum_k Q_k v^i \delta(\bm{x}-\bm{x}_k),
\quad (i=x,y,z),
\end{equation}
in the temporal gauge $A^0=0$.
This set of equations reproduces the ``hard thermal loop'' effective
theory~\cite{ClTransp} near equilibrium.
In the following, we assume that the fields only depend on time and on
one spatial coordinate, $x$, which reduces the Yang-Mills equations to
1+1 dimensions. The particles are allowed to propagate in three
spatial dimensions. This is referred to as 1D+3V simulations.

Numerical techniques to solve the classical field equations coupled to
colored point-particles have been developed in Ref.~\cite{HuMullerMoore}.
Our update algorithm is closely related to the one explained there,
which we briefly summarize.  We employ the so-called Nearest-Grid-Point
(NGP) method which simply counts the number of particles $N(j)$ within
a distance $\pm a/2$ of the $j$th lattice site to obtain the density
$n_j=N(j)/a$ (with $a$ the lattice spacing).  If a particle crosses a
cell, a current $J_x$ is generated.  For example, if a particle
crosses from site $i$ to $i+1$,
\begin{equation}
J_x(t,i)= \frac{gQ}{a^3 N_{\mathrm{test}}}
    \delta(\frac{t}{a}-\frac{t_{\mathrm{cross}}}{a})~.
\end{equation}
The color charge then has to be parallel transported to the next site,
\begin{equation}
  Q(i+1) = U_x^\dagger(i)Q(i)U_x(i)~.
\end{equation}
The gauge links are related to the continuum fields via $U_x(i)
=\exp(igaA_x(i))$. 
In this way, the continuity equation for color charge
is satisfied locally, together with Gauss's law.
At $t_{\mathrm{cross}}$ we also update the particle momentum $p_x$
by imposing energy-momentum conservation in the presence of
the chromo-electric field $E_x(i)$. On the other hand, the rotation of
a particle's momentum due to
the color-magnetic field is updated in every time step.

For 1D+3V simulations a major simplification arises from the fact that
the {\em transverse} current
\begin{equation}
 \bm{J}_\perp(t,\bm{x}) = \frac{g}{N_{\mathrm{test}}}
         \sum_i Q_i \bm{v}_\perp \delta(\bm{x}-\bm{x}_i(t))~,
\end{equation}
can be updated continuously in time. Note that the color
rotation due to the gauge fields $A_y$ and $A_z$
(which in fact become adjoint scalars in 1D) is also continuous in time.
Therefore, our transverse current is very smooth and much less noisy
than $J_x$ which is obtained in the impulse approximation.
However, for 1D-3V simulations the longitudinal current can also be
made sufficiently smooth by employing a large number of
test-particles. Such a ``brute-force'' approach is no longer feasible
for 3D-3V simulations.

\begin{figure}[htb]
\insertplot{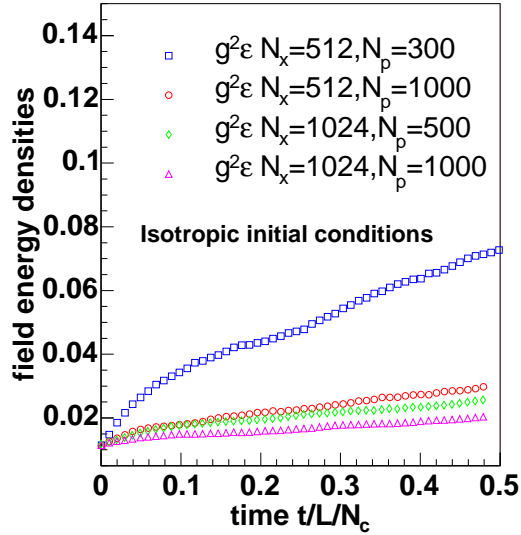}
\vspace*{-1cm}
\caption{
Time evolution of the average field energy density for {\em isotropic}
particle momentum distributions on two different lattices (with
the same physical size $L$) and for varying number of test
particles. $N_c=2$ color simulations.
}
\label{fig:fieldIso}
\end{figure}
To check the numerical accuracy, we have first performed simulations
for isotropic momentum distributions, varying the lattice spacing and
the number of test-particles (Fig.~\ref{fig:fieldIso}). The field
energy density is determined from the lattice field strength
$\bm{E}_L\equiv g a^2 \bm{E}$ as $g^2\epsilon = (1/2) \bm{E}_L^2/a^4$,
plus the magnetic contribution. One observes
that the time evolution stabilizes with increasing number of
test-particles and decreasing lattice spacing. However, to reach
sufficient accuracy our runs required several hundred to a thousand
test-particles per lattice site, corresponding to $\approx6$ hours
run-time on a single-processor 2.4 GHz Opteron workstation per initial
condition. It is therefore clear that a simple-minded extension of
the point-particle algorithm to 3D-3V simulations is impossible.
For multi-dimensional simulations, a generalization of
current smearing from U(1)~\cite{AbelianSmearing} to non-Abelian gauge
groups is essential. Non-Abelian simulations which include fluctuations of the
fields in the transverse plane would be very interesting because of
indications that this leads to the development of a turbulent cascade
which transfers energy from the soft unstable modes to stable UV field
modes (near $k\sim1/a$). This process, which is due to the self-interaction
of the gauge field in the SU(2) theory, effectively tames the
exponential growth of the fields~\cite{AMY}.

\section{Anisotropic initial distribution}

In what follows, we consider anisotropic initial momentum
distributions of the hard gluons,
\begin{equation}
  f(\bm{p},\bm{x})\propto \exp(-\sqrt{p_y^2+p_z^2}/p_\mathrm{hard})\,
\delta(p_x)~.
\end{equation}
This represents a quasi-thermal distribution in two dimensions, with
``temperature'' $=p_\mathrm{hard}$. 

The initial field amplitudes are sampled from a Gaussian distribution
with a width tuned to a given initial energy density.  We solve the
Yang-Mills equations in $A^0=0$ gauge and also set $\bm{A}=0$ at time
$t=0$; the initial electric field is taken to be polarized in a random
direction transverse to the $x$-axis. This initial condition is
convenient because Gauss's law $D_iE^i=\rho$ then implies
local (color) charge neutrality at $t=0$, allowing for a
straightforward initialization. Of course, magnetic and longitudinal
electric field components quickly build up as time progresses.

\begin{figure}[htb]
\insertplot{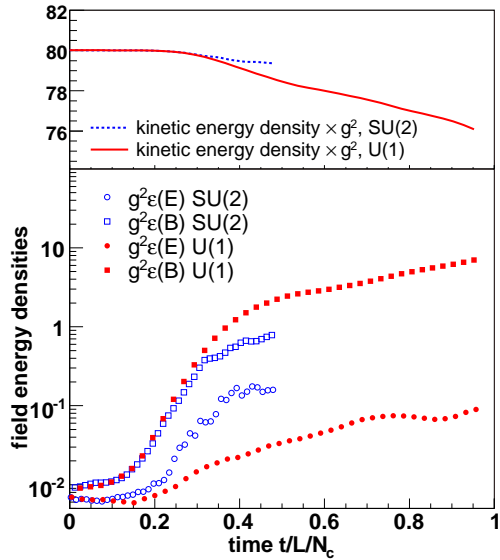}
\vspace*{-1cm}
\caption{
Time evolution of the kinetic (particle), and magnetic and electric
field energy densities for $U(1)$ and $SU(2)$ gauge group,
respectively.}
\label{fig:fieldW}
\end{figure}
We first show results for a relatively large separation of initial
particle and field energy densities which should qualitatively
resemble the conditions studied in~\cite{HL,AMY}. The results shown in
Fig.~\ref{fig:fieldW} correspond to a lattice of physical size
$L=40$~fm and $N_x=1024$ sites (the plots shown in ref.~\cite{DN05}
correspond to the same $L$ but half the number of lattice sites). The
hard scale was chosen as $p_\mathrm{hard}=10$~GeV (in lattice units,
$p_{L,hard}=ap_{hard}$), and the particle density
$g^2n=10$~fm$^{-3}$ (we define the density in lattice units as
$n_L=g^2 a^3 n$). The above definitions of the lattice hamiltonian,
fields and phase-space distribution function remove any explicit
reference to the gauge coupling $g$ from the lattice theory.

For the Abelian theory we observe a rapid exponential growth of the
magnetic field energy density, starting at about $t/L\approx 0.1$; we
repeat that in order to avoid a ``fake'' growth of the fields during
this initial transient time, one has to ensure that the number of
test particles is sufficiently large. At a time
$t/L\approx 0.4$ the magnetic field strength has grown by about one order of
magnitude. The fields clearly affect the particles, which
loose energy. In turn, at this time the exponential growth of
the magnetic fields is slowed down. The electric field grows less rapidly
and equipartitioning is not achieved within the depicted time
interval. 

The non-Abelian case features a rather similar evolution for short
times ($t/L/N_c\approx0.2$ for electric fields and $\approx0.3$ for
magnetic fields, respectively). We scaled time by $1/N_c$ because such
a scaling is natural in the linear regime~\cite{HL}. The
growth of the magnetic field then saturates somewhat earlier than for
the U(1) theory, and due to commutators, the electric field
has more strength by the end of the simulation. Due to the
somewhat earlier saturation of the instability, the colored particles
loose less of their energy to the fields than was the case for
electric charges. Nevertheless, at a purely qualitative level the
$U(1)$ and $SU(2)$ simulations are not extremely different, which is
due to the phenomenon of ``Abelianization'' in 1D-3V
simulations~\cite{Stan_review,HL}. This does not occur in 3D.

\begin{figure}[htb]
\insertplot{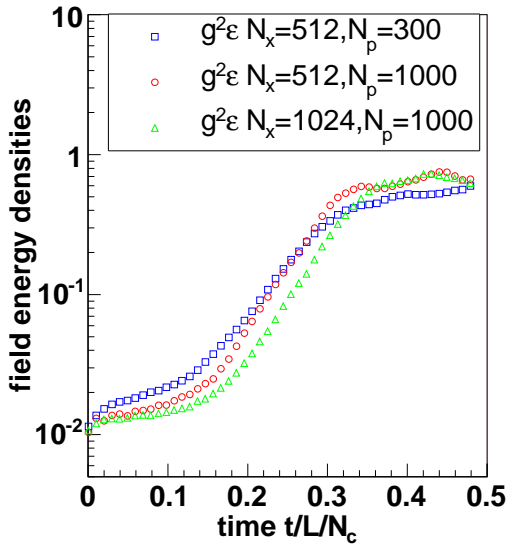}
\caption{Time evolution of the field energy density on
two different lattices (with the same
physical size $L$); $N_p$ denotes the number of test particles
per lattice site.}
\label{fig:LatComp}
\end{figure}
In Fig.~\ref{fig:LatComp} we compare results obtained on two different
lattices with $N_x=512$ and $N_x=1024$ sites, respectively, for the
same set of physical parameters. One observes that the growth rate of
the instability, the saturation level and time are nearly the
same. This confirms the underlying physical picture that the dynamics
is dominated by the unstable soft field modes rather than UV-modes
near the end of the Brioullin zone. If field modes with $k\sim1/a$
would affect the dynamics then the continuum limit would not exist.

Our 1D-3V simulations thus clearly confirm the existence of
instabilities in non-Abelian plasmas and the idea of
``Abelianization'', namely that the field growth is perhaps damped by
self-interactions but does not shut down until the fields have grown
so much as to affect the motion of the particles. Nevertheless, 
the number of $e$-foldings by which the field energy density grows is
much less spectacular in our simulations than for simulations within
the ``hard-loop'' approximation~\cite{HL}. This is due to the fact
that our initial field amplitudes are already relatively large
(non-linear regime). At a technical level, point-particle
simulations are not very well suited to study the extreme weak-field
regime, which would require a prohibitively large number of particles.

\begin{figure}[htb]
\begin{minipage}[t]{190bp}
\includegraphics[width=2.2in]{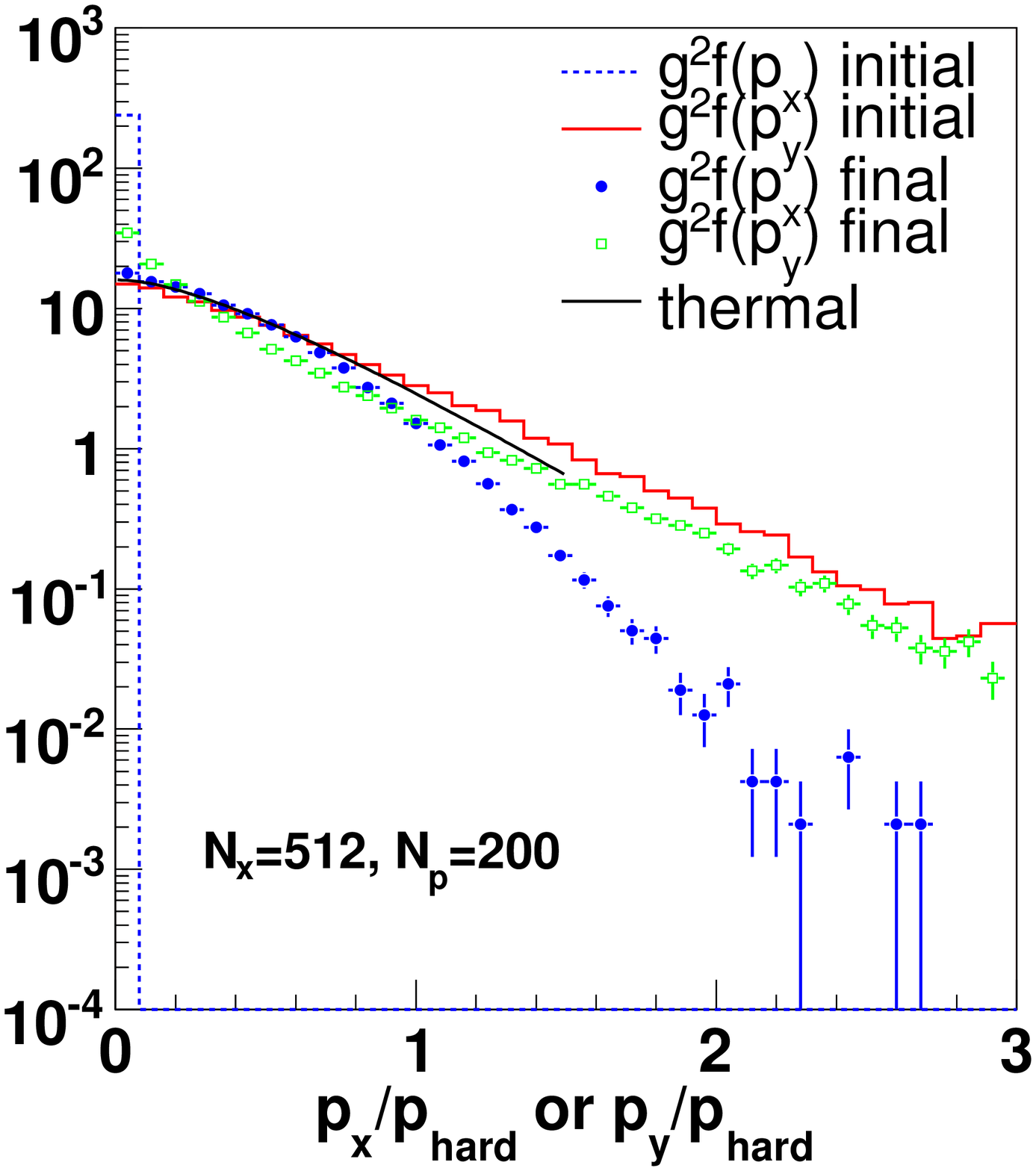}
\vspace*{-.6cm}
\caption{Initial and final particle distribution functions for the
  strong field case on a $N_x=512$ lattice.}
\label{fig:pts512}
\end{minipage}
\hspace{\fill}
\begin{minipage}[t]{190bp}
\includegraphics[width=2.2in]{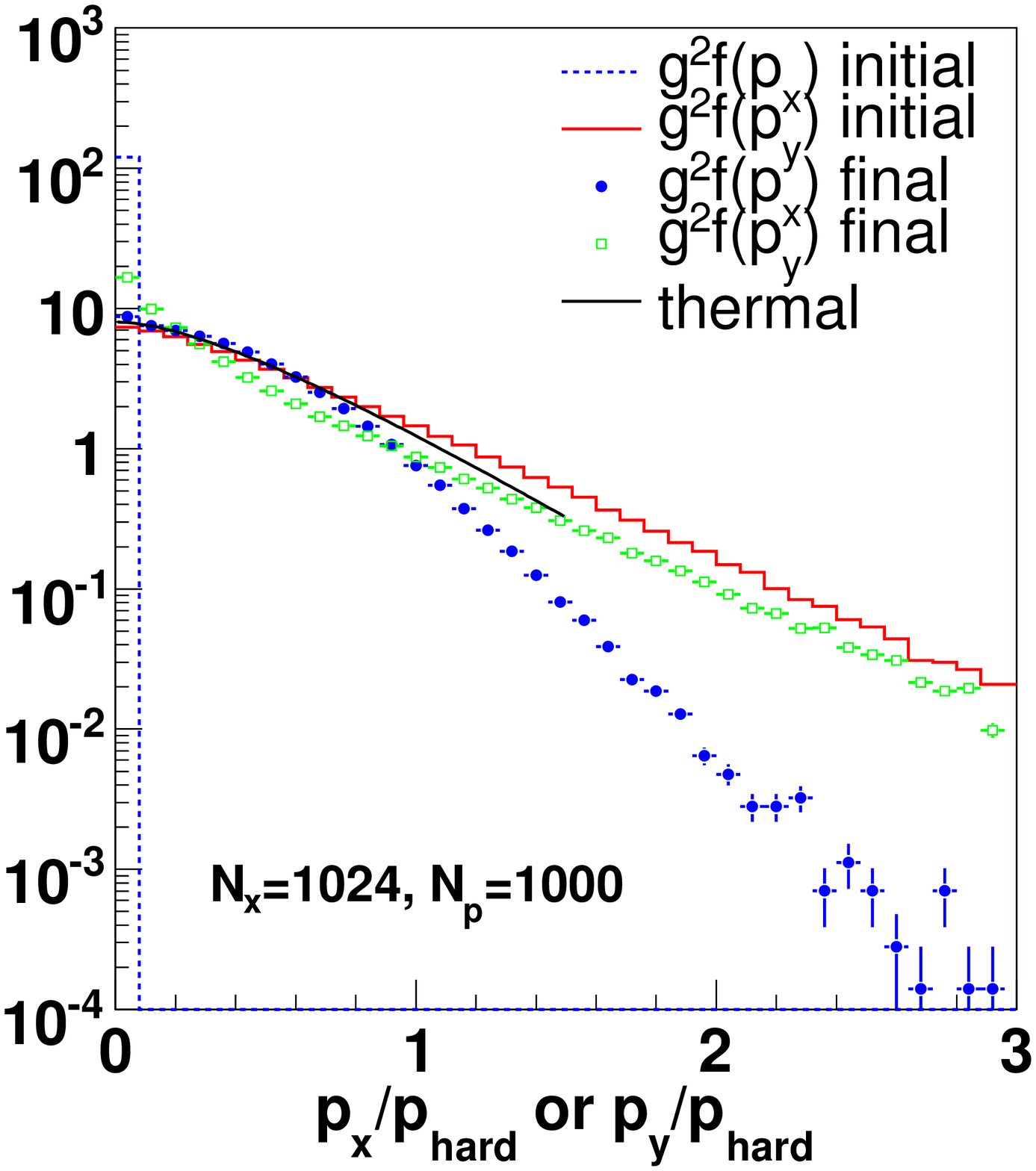}
\vspace*{-0.6cm}
\caption{Same as Fig.~\protect\ref{fig:pts512} for a $N_x=1024$ lattice.}
\label{fig:pts1024}
\end{minipage}\end{figure}
Once the fields have grown strong, they deflect the particles from
their straight-line trajectories and finally lead to isotropization of
their momentum distributions, shown explicitly in refs.~\cite{DN05}.
This process represents the dominant contribution to the build-up of
longitudinal pressure. Figs.~\ref{fig:pts512},\ref{fig:pts1024} depict
the evolution of the particle distribution function. This result was
obtained with ``strong-field'' initial conditions:
$p_{\mathrm{hard}}=1$~GeV and initial field energy density $\approx
10^{-1}$ GeV/fm$^3/g^2$; the time scale is set by the lattice size,
$L=10$~fm for this simulation.  When the separation between hard and
soft modes is not so large, strong instabilities can not develop as
the system approaches isotropy very quickly~\cite{DN05}.

In fact, propagation in strong random fields not only leads to
isotropic but even to nearly exponential particle momentum distributions,
as can be seen from Fig.~\ref{fig:pts512}. All particles with momenta up
to $\sim p_{hard}$ appear to be more or less thermalized. Once again,
we check the dependence on the lattice spacing by comparing results
obtained on different lattices. We confirm numerically that the
process does not appear to be dominated by the ultraviolet modes of
the fields on the lattice.

Field fluctuations generate an effective collision term
mediating soft exchanges.  Defining
\begin{equation}
 f(x,p,Q) = \langle f \rangle + \delta f, \quad
 A^a_\mu = \langle A^a_\mu \rangle + \delta A^a_\mu,
\end{equation}
where $\langle \ \rangle$ denotes the ensemble average, and
$\langle\delta f\rangle = \langle\delta A^a_\mu\rangle = 0$, one can
obtain the Balescu-Lenard collision term from the fluctuation part,
showing the correspondence between fluctuations and collisions in
an Abelian plasma~\cite{PhysicalKinetics}.  For the
non-Abelian case, see refs.~\cite{flucs}.  We also note that the mean
entropy density is no longer conserved in the presence of
fluctuations.

\section*{Acknowledgments}
We thank the organizers of the QGPTH05 workshop for the opportunity to
participate and to present our work, and, of course, for the invitation
to visit Vienna.


\begin{thebibliography}{99}  
  
\bibitem{QGPTH05}
{\sl QGPTH05}, Workshop on Quark-Gluon-Plasma Thermalization, August
10 -- 12, 2005, TU Wien, Austria;
\begin{verbatim}http://hep.itp.tuwien.ac.at/qgpth05\end{verbatim}

\bibitem{Stan_review}
S.~Mrowczynski,
``Instabilities driven equilibration of the quark-gluon plasma,''
arXiv:hep-ph/0511052, and references therein.

\bibitem{HL}
P.~Romatschke and M.~Strickland,
Phys.\ Rev.\ D {\bf 68}, 036004 (2003);
Phys.\ Rev.\ D {\bf 70}, 116006 (2004);\\
S.~Mrowczynski, A.~Rebhan and M.~Strickland,
Phys.\ Rev.\ D {\bf 70}, 025004 (2004);\\
P.~Arnold, J.~Lenaghan and G.~D.~Moore, JHEP {\bf 0308}, 002 (2003);\\
P.~Arnold and J.~Lenaghan, Phys.\ Rev.\ D {\bf 70}, 114007 (2004);\\
P.~Arnold, J.~Lenaghan, G.~D.~Moore and L.~G.~Yaffe,
Phys.\ Rev.\ Lett.\ {\bf 94}, 072302 (2005);\\
A.~Rebhan, P.~Romatschke and M.~Strickland,
Phys.\ Rev.\ Lett.\  {\bf 94}, 102303 (2005);\\
P.~Romatschke and R.~Venugopalan, arXiv:hep-ph/0510121.

\bibitem{DN05}
A.~Dumitru and Y.~Nara, Phys.\ Lett.\ B {\bf 621}, 89 (2005);\\
Y.~Nara, nucl-th/0509052.

\bibitem{Wong}
S.~K.~Wong, Nuovo Cim.\ A {\bf 65}, 689 (1970);\\
U.~W.~Heinz, Nucl.\ Phys.\ A {\bf 418} (1984) 603C.

\bibitem{ClTransp}
P.~F.~Kelly, Q.~Liu, C.~Lucchesi and C.~Manuel,
Phys.\ Rev.\ D {\bf 50}, 4209 (1994);\\
J.~P.~Blaizot and E.~Iancu,
Phys.\ Rept.\  {\bf 359}, 355 (2002).

\bibitem{HuMullerMoore}
C.~R.~Hu and B.~M\"uller,
Phys.\ Lett.\ B {\bf 409}, 377 (1997);\\
G.~D.~Moore, C.~R.~Hu and B.~M\"uller,
Phys.\ Rev.\ D {\bf 58}, 045001 (1998).

\bibitem{AbelianSmearing}
J.~Villasenor and O.~Buneman, Comp.\ Phys.\ Comm.\ {\bf 69}, 306 (1992).

\bibitem{AMY}
P.~Arnold, G.~D.~Moore and L.~G.~Yaffe,
Phys.\ Rev.\ D {\bf 72}, 054003 (2005);\\
A.~Rebhan, P.~Romatschke and M.~Strickland,
JHEP {\bf 0509}, 041 (2005);\\
P.~Arnold and G.~D.~Moore, 
arXiv:hep-ph/0509206;
arXiv:hep-ph/0509226.

\bibitem{PhysicalKinetics}
 E.~M.~Lifshitz and L.~P.~Pitaevskii, 
 Landau and Lifshitz Course of Theoretical Physics Vol.~10:
 ``Physical Kinetics''; Pergamon Press, Oxford, 1981.

\bibitem{flucs}
D.~F.~Litim and C.~Manuel,
Phys.\ Rev.\ Lett.\  {\bf 82}, 4981 (1999);
Nucl.\ Phys.\ B {\bf 562}, 237 (1999);\\
D.~B\"odeker, Nucl.\ Phys.\ B {\bf 559} (1999) 502.

\end{thebibliography}
\end{document}